\documentclass[12pt,halfline,a4paper]{ouparticle}
\newtheorem{theorem}{Theorem}[section]

\begin{document}

\title{On Nirmala indices–based entropy measures for the complex structure of ruthenium bipyridine}

\author{%
\name{H. M. Nagesh$^{1, *}$}
\address{Department of Science and Humanities, \\
PES University, Bangalore, India.} 
\\
\name{Muhammad Kamran Siddiqui$^{2}$}
\address{$^{2}$Department of Mathematics, COMSATS University Islamabad,\\
Lahore Campus, Lahore, Pakistan.}
\email{email: kamransiddiqui75@gmail.com$^{2}$.} \\
}

\abstract{A numerical parameter, known as a topological index, is employed to represent the molecular structure of a compound by considering its graph-theoretical properties. In the study of quantitative structure-activity relationships (QSAR) and quantitative structure-property relationships (QSPR), topological indices are used to predict the physicochemical properties of chemical compounds. Graph entropies have evolved as information-theoretic tools to investigate the structural information of a molecular graph. In this study, we compute the Nirmala index, the first and second inverse Nirmala index of the complex structure of ruthenium bipyridine, with the help of its M-polynomial. Furthermore, entropy measures for the complex structure of ruthenium bipyridine are computed using Shannon's entropy model. The comparison between the Nirmala indices and their associated entropy measures is presented through numerical computation. The correlation between the Nirmala indices and associated entropy measurements is then examined using a regression model. 
}

\date{}

\maketitle
\newpage

\section{Introduction}
\label{sec1} 
Topological indices are structural invariants based on molecular graphs that capture the underlying connectivity of molecular networks, and thus these indices have received considerable attention in recent years owing to their applications in quantitative structure-activity and quantitative structure-property relationships (QSPR) relations \cite{1,2,3,4}. Degree-based topological indices have been studied extensively and they have been employed to predict the physicochemical properties of molecular structures \cite{5}. Information entropy metrics can be used to characterize the information complexity of complex chemical structures, including the complex structure of ruthenium bipyridine. Shannon first proposed the idea of information entropy to assess and measure the complexity of data and information transmission, but it has since been extensively used in many different scientific domains. The analysis of molecular structure complexity and quantum chemical electron densities [6] is one of the most significant uses of information entropy. In QSAR and QSPR investigations, topological indices combined with entropy measures may be a more effective tool. Information entropy has been discovered to be directly correlated with the physical features of fullerenes in several kinds of natural substances, including their formal carbon atom oxidation states and rotational symmetry numbers [7].

Let $\Upsilon =(V(\Upsilon), E(\Upsilon))$ be an ordered pair of a simple, connected, and undirected graph with non-empty vertex set $V(\Upsilon)$ and edge set $E(\Upsilon)$. The total number of edges incident to a vertex $v \in V(\Upsilon)$ is known as $degree$ of the vertex $v$ and is denoted as $d_{\Upsilon}(v)$. Let $e=pq$ represent an edge of the graph $\Upsilon$, where $p$ and $q$ are the end vertices of the edge $e$. The topological index based on degrees, defined on the edge set $E(\Upsilon)$ of a graph $\Upsilon$ \cite{8} is given by
    \begin{equation*}
I(\Upsilon)= \displaystyle \sum_{uv \in E(\Upsilon)} f(d_{\Upsilon}(u), d_{\Upsilon}(v)),
    \end{equation*}
where $f(x, y)$ is a non-negative and symmetric function that depends on the mathematical formulation of the topological index.

Several topological indices have been reported in the literature and have shown benefits in several fields, including drug development, biology, chemistry, computer science, and physics. Proposed by H. Wiener in 1947, the Wiener index was the first and most studied topological index. One noteworthy application is the prediction of paraffin boiling temperatures \cite{9}. Another well-known degree-based topological statistic is the connectedness index, sometimes known as the \emph{Randi\'c index} and first introduced by Milan Randi\'c in 1975. Its importance in making drugs is generally recognized \cite{10}. For additional information on topological indices and their applications, readers are referred to \cite{11,12}. 

Numerous efforts have been undertaken to improve the category of degree-based topological indices by the addition of new indices. \newpage Kulli in \cite{13} introduced a novel degree-based topological index of a molecular graph $\Upsilon$, which is called the \emph{Nirmala index} as follows.

\begin{equation}
N(\Upsilon)= \displaystyle \sum_{uv \in E(\Upsilon)} \sqrt{d_{\Upsilon}(u)+d_{\Upsilon}(v)}
    \end{equation}
  
Later in 2021, Kulli \cite{14} introduced the notion of the first inverse Nirmala index $IN_{1}(\Upsilon)$ and second inverse Nirmala index $IN_{2}(\Upsilon)$ of a molecular graph $\Upsilon$ as follows.
\begin{equation}
IN_{1}(\Upsilon)= \displaystyle \sum_{uv \in E(\Upsilon)} \sqrt{\frac{1}{d_{\Upsilon}(u)}+\frac{1}{d_{\Upsilon}(v)}}=\displaystyle \sum_{uv \in E(\Upsilon)} \left( \frac{1}{d_{\Upsilon}(u)}+\frac{1}{d_{\Upsilon}(v)} \right)^{\frac{1}{2}}
    \end{equation}
 \begin{equation}
IN_{2}(\Upsilon)= \displaystyle \sum_{uv \in E(\Upsilon)} \frac{1}{\sqrt{\frac{1}{d_{\Upsilon}(u)}+\frac{1}{d_{\Upsilon}(v)}}}=\displaystyle \sum_{uv \in E(\Upsilon)} \left( \frac{1}{d_{\Upsilon}(u)}+\frac{1}{d_{\Upsilon}(v)} \right)^{-\frac{1}{2}}
    \end{equation} 
    
 \vspace{5mm}   
In the past, several topological indices were computed utilizing their conventional mathematical formulation. There are several attempts to look into a compact method that can recover many topological indices of a particular class. In this regard, the concept of a general polynomial was developed, whose values of the necessary topological indices at a given point are produced by its derivatives, integrals, or a combination of both. For instance, the distance-based topological indices are recovered by the Hosoya polynomial \cite{15}, whereas the NM-polynomial generates the neighborhood degree sum-based topological indices \cite{16}. 

The M-polynomial was introduced by Deutsch and Kla\v{z}ar in \cite{17} to find the degree-based topological indices.  \\\\
\textbf{Definition 1.1} (\cite{17}) The M-polynomial of a graph $\Upsilon$ is defined as:
\begin{center}
    $M(\Upsilon; x,y)=\displaystyle \sum_{\delta \leq i \leq j \leq \Delta} m_{i,j}(\Upsilon)x^{i}y^{j}$,
\end{center}
where $\delta = min\{d_{\Upsilon}(u) | u \in V(\Upsilon)\}$, $\Delta = max\{d_{\Upsilon}(u) | u \in V(\Upsilon)\}$, and $m_{ij}$ is the number of edges $uv \in E(\Upsilon)$ such that $d_{\Upsilon}(u)=i, d_{\Upsilon}(v)=j \, (i,j \geq 1)$.

The M-polynomial-based derivation formulae to compute the different Nirmala indices are listed in Table 1.
\newpage
\begin{center}
 \textbf{Table 1}. Relationship between the M-polynomial and Nirmala indices for a graph $\Upsilon$.      
\end{center}

\begin{table}[h!]
\centering
\renewcommand{\arraystretch}{2.5}
\begin{tabular}{||c| c |c| c ||} 
 \hline
 Sl. No & Topological Index & $f(x,y)$ & Derivation from $M(\Upsilon;x,y)$  
 \\ [0.5ex] 
 \hline\hline
 1  & Nirmala index (N) & $\sqrt(x+y)$ & $D_{x}^{\frac{1}{2}}J(M(\Upsilon; x,y))|_{x=1}$ \\
\hline
2  & First inverse Nirmala index $(IN_1)$ & $\sqrt(\frac{x+y}{xy})$ & $D_{x}^{\frac{1}{2}}JS_{y}^{\frac{1}{2}}S_{x}^{\frac{1}{2}}(M(\Upsilon; x,y))|_{x=1}$\\
\hline
 3  & Second inverse Nirmala index $(IN_2)$ & $\sqrt(\frac{xy}{x+y})$ & $S_{x}^{\frac{1}{2}}JD_{y}^{\frac{1}{2}}D_{x}^{\frac{1}{2}}(M(\Upsilon; x,y))|_{x=1}$  \\
 [1ex] 
 \hline
\end{tabular}
\end{table} 
\vspace{5mm}
Here, $D_{x}^{\frac{1}{2}}(h(x,y))=\sqrt{x \cdot \frac{\partial(h(x,y))}{\partial x}} \cdot \sqrt{h(x,y)}$; \\ $D_{y}^{\frac{1}{2}}(h(x,y))=\sqrt{y \cdot \frac{\partial(h(x,y))}{\partial y}} \cdot \sqrt{h(x,y)}$;\\
$S_{x}^{\frac{1}{2}}(h(x,y))=\sqrt{\displaystyle \int_{0}^{x} \frac{h(t,y)}{t} }dt \cdot \sqrt{h(x,y)}$;\\
$S_{y}^{\frac{1}{2}}(h(x,y))=\sqrt{\displaystyle \int_{0}^{y} \frac{h(x,t)}{t} }dt \cdot \sqrt{h(x,y)}$; and $J(h(x,y))=h(x,x)$ are the operators. \\

We refer readers to \cite{18,19,20,21,22,23,24} for additional information on degree-based topological indices utilizing the M-polynomial.

Shannon \cite{25} first proposed the concept of entropy, which is a probability distribution-based measure of the uncertainty or unpredictability of the information in a system. Next, entropy was used to assess the structural information of networks, graphs, and chemical structures. In the past few years, graph entropies have gained increased utility across various fields, including mathematics, computer science, biology, chemistry, sociology, and ecology. There are various forms of graph entropy measurements, including extrinsic and intrinsic measures. These measures relate to the probability distribution concerning graph invariants, such as edges and vertices. Interested readers are referred to \cite{26,27,28,29} for additional details on degree-based graph entropy metrics and their applications.

\subsection{Entropy of a graph in terms of vertex degree}
Let $\Upsilon$ be a simple and connected graph with order $p$ and size $q$. Then, the graph entropy of a graph $\Upsilon$ as given in \cite{25} is defined as follows.
\begin{equation}
ENT_{\omega}(\Upsilon)=-\displaystyle \sum_{i=1}^{p}  \frac{\omega(s_{i})}{\displaystyle \sum_{j=1}^{p} \omega(s_j) } 
log \left(\frac{\omega(s_{i})}{\displaystyle \sum_{j=1}^{p} \omega(s_j) }  \right)
\end{equation} 
Here, $\omega$ is a meaningful information function and $s_i \in V(\Upsilon)$ for every $i \in \left[1,2,\ldots, p \right]$.

Let $\omega(s_i)=d_{\Upsilon}(s_i)$. Then equation (4) reduces to 

\begin{align*}
ENT_{\omega}(\Upsilon)=&-\displaystyle \sum_{i=1}^{p}  \frac{d_{\Upsilon}(s_i)}{\displaystyle \sum_{j=1}^{p} d_{\Upsilon}(s_j) } 
log \left(\frac{d_{\Upsilon}(s_i)}{\displaystyle \sum_{j=1}^{p} d_{\Upsilon}(s_j) }  \right) \\
=& log \left(\displaystyle \sum_{j=1}^{p} d_{\Upsilon}(s_j) \right)-\frac{1}{\displaystyle \sum_{j=1}^{p} d_{\Upsilon}(s_j)}     \displaystyle \sum_{i=1}^{p} d_{\Upsilon}(s_i) 
log(d_{\Upsilon}(s_i)) 
\end{align*}
By the fundamental theorem of graph theory, $\displaystyle \sum_{i=1}^{p} d_{\Upsilon}(s_i)=2q$.\\
Hence, 
\begin{align}
ENT_{\omega}(\Upsilon)=& log(2q)-\frac{1}{2q}     \displaystyle \sum_{i=1}^{p} d_{\Upsilon}(s_i) log(d_{\Upsilon}(s_i))
\end{align} 

\subsection{Entropy of a graph in terms of edge-weight}
 In 2014, Chen et al. \cite{30} proposed the concept of the entropy of edge-weighted graphs as follows. 
 
 Let $\Upsilon =(V(\Upsilon), E(\Upsilon), \omega(st))$ be an edge-weight graph, where $V(\Upsilon)$ is a set of vertices,  $E(\Upsilon)$ is a set of edges; and $\omega(st)$ denotes the weight of an edge  $st \in E(\Upsilon)$. Then the entropy of a graph in terms of edge weight is defined as, \vspace{3mm} 
\begin{align*}
ENT_{\omega}(\Upsilon)=&-\displaystyle \sum_{s^{'}t^{'} \in E(\Upsilon)} \frac{\omega(s^{'}t^{'})}{\displaystyle \sum_{st \in E(\Upsilon) } \omega(st) } 
log \left(\frac{\omega(s^{'}t^{'})}{\displaystyle \sum_{st \in E(\Upsilon) } \omega(st) } \right) \\
= & -\displaystyle \sum_{s^{'}t^{'} \in E(\Upsilon)} \frac{\omega(s^{'}t^{'})}{\displaystyle \sum_{st \in E(\Upsilon) } \omega(st) } 
\left[log(\omega(s^{'}t^{'})) -log \left(\displaystyle \sum_{st \in E(\Upsilon)} \omega(st) \right) \right] \\
=& log \left(\displaystyle \sum_{st \in E(\Upsilon)} \omega(st) \right) -\displaystyle \sum_{s^{'}t^{'} \in E(\Upsilon)} \frac{\omega(s^{'}t^{'})}{\displaystyle \sum_{st \in E(\Upsilon) } \omega(st) } 
log(\omega(s^{'}t^{'})) 
\end{align*}
Hence,
\begin{align}
ENT_{\omega}(\Upsilon)=& log \left(\displaystyle \sum_{st \in E(\Upsilon)} \omega(st) \right) - \frac{1}{\left(\displaystyle \sum_{st \in E(\Upsilon) } \omega(st) \right)}              \displaystyle \sum_{s^{'}t^{'} \in E(\Upsilon)} \omega(s^{'}t^{'})  log(\omega(s^{'}t^{'}))
\end{align} 

In 2023, Virendra Kumar et al. \cite{31} introduced the notion of the Nirmala indices-based entropy by considering the meaningful information function $\omega$ as a function associated with the definitions of the Nirmala indices as given in equations (1-3).
 \\\\
\textbf{Nirmala entropy}: Let $\omega(st)=\sqrt{d_{\Upsilon}(s)+d_{\Upsilon}(t)}$. Then from the definition of the Nirmala index as given in Equation (1), we have
\begin{center}
$ \displaystyle \sum_{st \in E(\Upsilon)}\omega(st)=\displaystyle \sum_{st \in E(\Upsilon)} \sqrt{d_{\Upsilon}(s)+d_{\Upsilon}(t)}=N(\Upsilon)$  
\end{center}
Hence, using equation (6), the Nirmala entropy of a graph $\Upsilon$ is given by 
\begin{equation}
 ENT_{N}(\Upsilon)=log(N(\Upsilon))  - \frac{1}{N(\Upsilon)}              \displaystyle \sum_{st \in E(\Upsilon)} \sqrt{d_{\Upsilon}(s)+d_{\Upsilon}(t)} \times log (\sqrt{d_{\Upsilon}(s)+d_{\Upsilon}(t)})
\end{equation} \\\\
\textbf{First inverse Nirmala entropy}: Let $\omega(st)=\sqrt{\frac{1}{d_{\Upsilon}(s)}+\frac{1}{d_{\Upsilon}(t)}}$. Then from the definition of the first inverse Nirmala index as given in Equation (2), we have,
\begin{center}
$ \displaystyle \sum_{st \in E(\Upsilon)}\omega(st)=\displaystyle \sum_{st \in E(\Upsilon)} \sqrt{\frac{1}{d_{\Upsilon}(s)}+\frac{1}{d_{\Upsilon}(t)}}=IN_{1}(\Upsilon)$  
\end{center}
Hence, using equation (6), the first inverse Nirmala
entropy of a graph $\Upsilon$ is given by 
\begin{equation}
 ENT_{IN_{1}}(\Upsilon)=log(IN_{1}(\Upsilon))  - \frac{1}{IN_{1}(\Upsilon)}              \displaystyle \sum_{st \in E(\Upsilon)} \sqrt{\frac{1}{d_{\Upsilon}(s)}+\frac{1}{d_{\Upsilon}(t)}} \times log \left( \sqrt{\frac{1}{d_{\Upsilon}(s)}+\frac{1}{d_{\Upsilon}(t)}} \right) 
\end{equation}
\textbf{Second inverse Nirmala entropy}: Let $\omega(st)=\frac{\sqrt{d_{\Upsilon}(s) \cdot  d_{\Upsilon}(t)} }{\sqrt{d_{\Upsilon}(s) + d_{\Upsilon}(t)} }$.
Then from the definition of the second inverse Nirmala index as given in Equation (3), we have,
\begin{center}
$ \displaystyle \sum_{st \in E(\Upsilon)}\omega(st)=\displaystyle \sum_{st \in E(\Upsilon)} \frac{\sqrt{d_{\Upsilon}(s) \cdot  d_{\Upsilon}(t)} }{\sqrt{d_{\Upsilon}(s) + d_{\Upsilon}(t)} }=IN_{2}(\Upsilon)$  
\end{center}
Hence, using equation (6), the second inverse Nirmala
entropy of a graph $\Upsilon$ is given by 
\begin{equation}
 ENT_{IN_{2}}(\Upsilon)=log(IN_{2}(\Upsilon))  - \frac{1}{IN_{2}(\Upsilon)}              \displaystyle \sum_{st \in E(\Upsilon)} \frac{\sqrt{d_{\Upsilon}(s) \cdot  d_{\Upsilon}(t)} }{\sqrt{d_{\Upsilon}(s) + d_{\Upsilon}(t)} } \times log \left( \frac{\sqrt{d_{\Upsilon}(s) \cdot  d_{\Upsilon}(t)} }{\sqrt{d_{\Upsilon}(s) + d_{\Upsilon}(t)} } \right) 
 \end{equation}
 
 \subsection{Methodology}
In this paper, the complicated structural details of ruthenium bipyridine are obtained. Further, we use Shannon's entropy measures by employing novel information functions derived from various Nirmala indices definitions. We conduct a comprehensive mathematical and computational exploration of these measures within the complex structure of ruthenium bipyridine. The rest of the paper is organized as follows: Section 2 discusses the 2D and 3D structure of ruthenium bipyridine. In Section 3, we compute the Nirmala indices of ruthenium bipyridine using its M-polynomial, which allows us to calculate the Nirmala indices-based entropy measures of ruthenium bipyridine $RB_n$. Section 4 compares the Nirmala indices and their associated entropy measures through numerical data. Section 5 deals with a regression model to illustrate how the estimated Nirmala indices and associated entropy values fit the curve. Lastly, a discussion and a conclusion are presented in sections 6 and 7, respectively.  

\section{Complex structure of ruthenium bipyridine}
Ruthenium (II) tris (bipyridine) is one of the compounds that has had the largest impact on the advancement of chemistry. The solvation of ruthenium dyes \cite{32} in various solvents and electrolytes has received a lot of interest, particularly in Dye-Sensitive Solar Cells (DSSCs). The emission and absorption spectra of ruthenium tris (bipyridine) complexes \cite{33} adorned with various electron-rich and redox-active amine substituents exhibit reversible changes in the deep red to near-infrared range when low electrochemical potentials are applied. \newpage Multiple targets and pathways are involved in the anticancer actions of ruthenium complexes. Comparing the ruthenium complexes to other metal-based molecules under study, they demonstrated impressive anticancer activity. Because of their great selectivity and low toxicity, organometallic ruthenium (II) complexes have been thoroughly investigated for possible biological applications. Several ruthenium compounds that effectively increased selectivity to various processes by strictly controlling the coordination sphere have been the subject of numerous investigations \cite{34,35}. For a ruthenium core, a range of molecular structures can be readily constructed because of ruthenium's versatility in oxidation states and its easy interaction with ligands. The readers are referred to \cite{36} for further information and an analysis of the properties and uses of ruthenium. 

This paper aims to support scientists in their exploration of ruthenium bipyridine's physical properties. Since ruthenium is among the rarest metals, conducting trials would be too expensive to influence the price. Therefore, cost management can be achieved by mathematical chemistry. It is far more time and money-consuming for a chemist to do lab tests to investigate the properties of the chemical than it is for a mathematician to research the chemical's structure, which is free and allows for quicker discovery of the qualities. Figure 1 shows the 2D and 3D structure of complex ruthenium bipyridine \cite{37}.
\vspace{5mm}
\begin{figure}[h!]
\centering
\includegraphics[width=200mm]{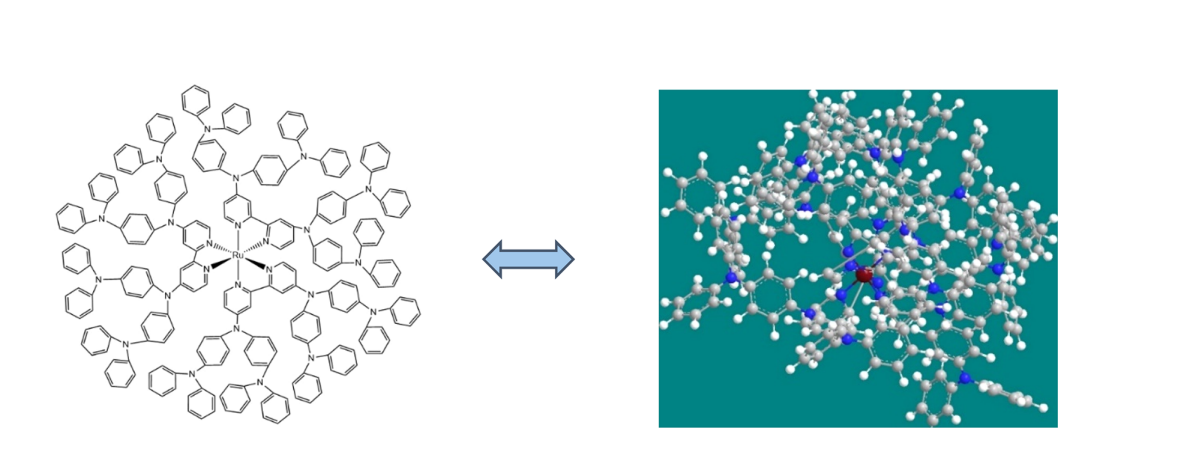}  
  \end{figure} 
  \begin{center}
  Figure 1. 2D and 3D structure of complex ruthenium bipyridine.    
  \end{center}
  \newpage
 \subsection{Complex structure of ruthenium bipyridine graph} 
 By incorporating triphenylamine into the coordination molecule of ruthenium tris bipyridine, the ruthenium bipyridine graph's complex structure is formed, with each atom—both metal and organic ligand—representing the graph's vertices and bonds acting as its edges. The molecular graph of the complex structure of ruthenium bipyridine $RB_n$ is shown in Figure 2, where $n$ is the number of triphenylamine polymers around the ruthenium tris bipyridine complexes. 
 The order and size of $RB_n$ are 
\begin{center}
    $|V(RB_n)=6 \left[6(2^n)+6(2^n-1)\right]+\left[6(2^{n-2})+6(2^{n-2}-1) \right]+1$ and \\
    $|E(RB_n)=6 \left[6(2^{n+1})+6(2^{n+1}-1)\right]+3\left[6(2^{n})+6(2^{n}-1) \right]+9$, respectively.
\end{center}

\begin{figure}[h!]
\centering
\includegraphics[width=210mm]{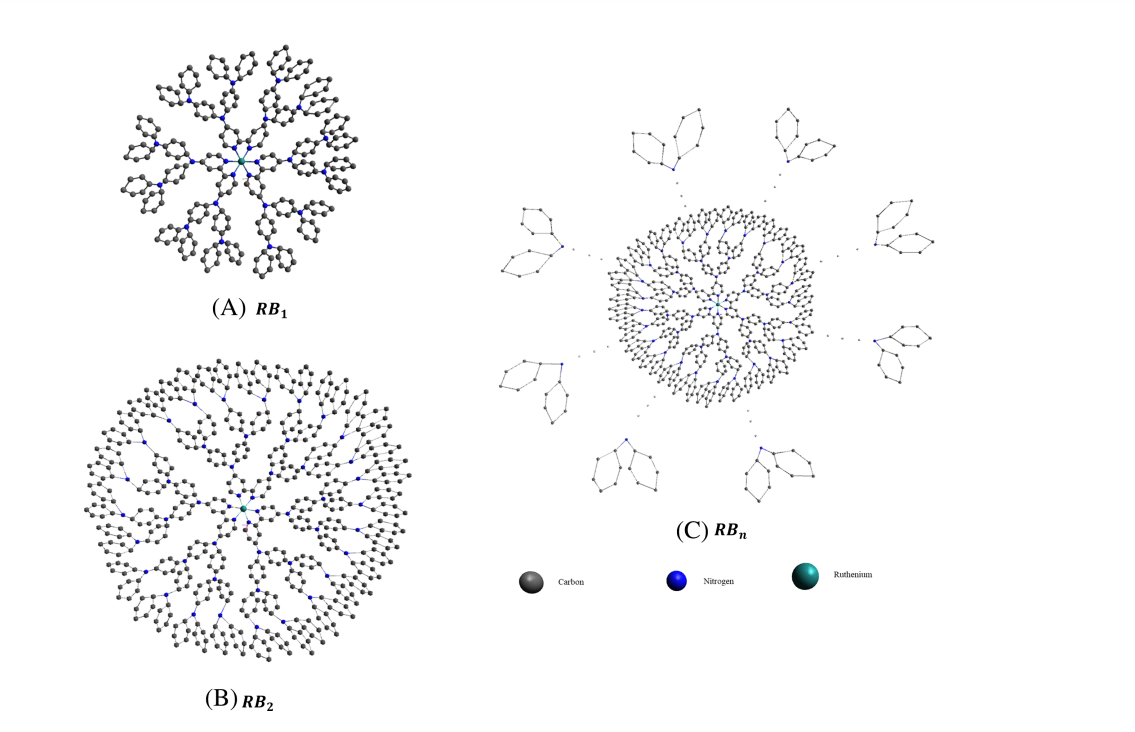}  
  \end{figure} 
  \begin{center}
  Figure 2. The complex structure of ruthenium bipyridine graph $RB_n$.    
  \end{center}
  
Motivated by the studies described in [31, 37], we aim to calculate the Nirmala index, as well as the first and second inverse Nirmala index, for the complex structure of ruthenium bipyridine $RB_n$ using M-polynomial.  Furthermore, entropy measures for the complex structure of ruthenium bipyridine are computed using Shannon's entropy model.

\newpage 
\section{Main results}

In this section, we first find the M-polynomial of ruthenium bipyridine $RB_n$. Then, we compute the Nirmala index, and first and second inverse Nirmala indices using its M-polynomial. For the complex structure of ruthenium bipyridine graph $RB_n$, the size of $RB_n$ is
\begin{center}
       $|E(RB_n)|=6 \left[6(2^{n+1})+6(2^{n+1}-1)\right]+3\left[6(2^{n})+6(2^{n}-1) \right]+9$.
\end{center}

The edge set partitions of the complex structure of ruthenium bipyridine graph $RB_n$ are shown in Table 2.
\vspace{5mm}
\begin{table}[h!]
\centering
\renewcommand{\arraystretch}{2.5}
\begin{tabular}{||c| c |c| c ||} 
 \hline
 Sl. No & Edge set & $(d_{\Upsilon}(r),d_{\Upsilon}(s))$ & Number of repetitions  
 \\ [0.5ex] 
 \hline\hline
 1  & $E_1$  & (2,2) & $18 \cdot 2^{n+2}-18$\\
\hline
2  & $E_2$ & (2,3) & $36 \cdot 2^{n+1}-24$ \\
\hline
 3  & $E_3$ & (3,3) & $9 \cdot 2^{n+2}-9$  \\
 \hline
 4  & $E_4$ & (3,6) & $6$  \\
 \hline 
\end{tabular}
\end{table} 
\begin{center}
    \textbf{Table 2}. Edge set partition of $RB_n$ according to degrees of end vertices of an edge. 
\end{center}
It is easy to observe that 
\begin{equation*}
|E(RB_n)|=\left[6(2^{n+1})+6(2^{n+1}-1)\right]+3\left[6(2^{n})+6(2^{n}-1) \right]+9=72 \cdot 2^{n+1}+36 \cdot 2^n-45.    
\end{equation*}
\begin{equation*} \displaystyle \sum_{i=1}^{4}|E_{i}|=27 \cdot 2^{n+2}+36 \cdot 2^{n+1}-45.  
\end{equation*}
Using the laws of exponents, one can easily prove that
\begin{equation*}
72 \cdot 2^{n+1}+36 \cdot 2^n-45=180 \cdot 2^n-45.    
\end{equation*}
\begin{equation*}
27 \cdot 2^{n+2}+36 \cdot 2^{n+1}-45=180 \cdot 2^n-45.   
\end{equation*}
Therefore, $|E(RB_n)|=\displaystyle \sum_{i=1}^{4}|E_{i}|$.

\subsection{Nirmala indices of $RB_n$}
We now find the M-polynomial of $RB_n$ as follows:

\begin{theorem}
Let $\Upsilon=RB_n$ be the molecular graph for the complex structure of ruthenium bipyridine. Then the M-polynomial of $\Upsilon$ is \\
$M(\Upsilon; x,y)=(18 \cdot 2^{n+2}-18)x^2y^2+(36 \cdot 2^{n+1}-24)x^2y^3+ (9 \cdot 2^{n+2}-9)x^3y^3+ 6x^3y^6$.
\end{theorem}
\textbf{Proof.} Let $\Upsilon=RB_n$ be the molecular graph for the complex structure of ruthenium bipyridine.\\ From Table 2,  $\displaystyle \sum_{i=1}^{4}|E_{i}|=27 \cdot 2^{n+2}+36 \cdot 2^{n+1}-45$. 

Since each vertex of $\Upsilon$ is of degree either $2$ or $3$ or $6$, 
the partitions of edge set $E(\Upsilon)$ are: \\
$E_{1}(\Upsilon):=\{e=ab \in E(\Upsilon): d_{\Upsilon}(a)=2, d_{\Upsilon}(b)=2 \}$;\\
$E_{2}(\Upsilon)):=\{e=ab \in E(\Upsilon): d_{\Upsilon}(a)=2, d_{\Upsilon}(b)=3 \}$;\\
$E_{3}(\Upsilon):=\{e=ab \in E(\Upsilon): d_{\Upsilon}(a)=3, d_{\Upsilon}(b)=3 \}$;\\
$E_{4}(\Upsilon):=\{e=ab \in E(\Upsilon): d_{\Upsilon}(a)=3, d_{\Upsilon}(b)=6 \}$.\\\\
Clearly, $|E_{1}(\Upsilon)|=18 \cdot 2^{n+2}-18; |E_{2}(\Upsilon)|=36 \cdot 2^{n+1}-24; |E_{3}(\Upsilon)|=9 \cdot 2^{n+2}-9;\\ |E_{4}(\Upsilon)|=6$. Therefore, 
\begin{align*}
M(\Upsilon; x,y)=&\displaystyle \sum_{\delta \leq i \leq j \leq \Delta} m_{i,j}(\Upsilon)x^{i}y^{j} = m_{22}(\Upsilon)x^{2}y^{2}+m_{23} (\Upsilon)x^{2}y^{3}+m_{33}(\Upsilon)x^{3}y^{6} + m_{36}(\Upsilon)x^{3}y^{6} \\ 
 = &(18 \cdot 2^{n+2}-18)x^2y^2+(36 \cdot 2^{n+1}-24)x^2y^3+ (9 \cdot 2^{n+2}-9)x^3y^3+ 6x^3y^6.
 \end{align*}
 
Now we evaluate the Nirmala indices of $RB_n$ with the help of its M-polynomial.

\begin{theorem}
Let $\Upsilon=RB_n$ be the molecular graph for the complex structure of ruthenium bipyridine. Then the Nirmala indices of $\Upsilon$ are:\\\\
a. $N(\Upsilon)=(36 \cdot 2^{n+2}-36)+\sqrt5(36 \cdot 2^{n+1}-24)+\sqrt6(9 \cdot 2^{n+2}-9)+18$,\\\\
b. $IN_{1}(\Upsilon)= (18 \cdot 2^{n+2}-18)+\frac{\sqrt{5}}{\sqrt{6}}(36 \cdot 2^{n+1}-24)+\frac{\sqrt{6}}{3}(9\cdot 2^{n+2}-9)+\frac{6}{\sqrt{2}}$,\\\\
c. $IN_{2}(\Upsilon)=(18 \cdot 2^{n+2}-18)+\frac{\sqrt{6}}{\sqrt{5}}(36 \cdot 2^{n+1}-24)+\frac{27}{\sqrt{6}}(2^{n+2}-1)+6\sqrt{2}$.
\end{theorem}
\textbf{Proof}. Let $\Upsilon=RB_n$ be the molecular graph for the complex structure of ruthenium bipyridine. From Theorem 3.1, the M-polynomial of $\Upsilon$ is
\begin{center}
$M(\Upsilon; x,y)=(18 \cdot 2^{n+2}-18)x^2y^2+(36 \cdot 2^{n+1}-24)x^2y^3+ (9 \cdot 2^{n+2}-9)x^3y^3+ 6x^3y^6$.
\end{center} 
From Table 2, we have 
\begin{align*}
& (i) \, D_{x}^{\frac{1}{2}}J(M(\Upsilon; x,y)) \\
& = D_{x}^{\frac{1}{2}}J \left[(18 \cdot 2^{n+2}-18)x^2y^2+(36 \cdot 2^{n+1}-24)x^2y^3+ (9 \cdot 2^{n+2}-9)x^3y^3+ 6x^3y^6 \right] \\
& = D_{x}^{\frac{1}{2}} \left[(18 \cdot 2^{n+2}-18)x^4+(36 \cdot 2^{n+1}-24)x^5+ (9 \cdot 2^{n+2}-9)x^6+ 6x^9 \right] \\
& =2(18 \cdot 2^{n+2}-18)x^4+\sqrt{5}(36 \cdot 2^{n+1}-24)x^5+ \sqrt{6}(9 \cdot 2^{n+2}-9)x^6+ 18x^9 \\
& =(36 \cdot 2^{n+2}-36)x^4+\sqrt{5}(36 \cdot 2^{n+1}-24)x^5+ \sqrt{6}(9 \cdot 2^{n+2}-9)x^6+ 18x^9 . 
\end{align*}
\begin{align*}
& (ii) \, D_{x}^{\frac{1}{2}}JS_{y}^{\frac{1}{2}}S_{x}^{\frac{1}{2}}(M(\Upsilon; x,y)) \\
=&D_{x}^{\frac{1}{2}}JS_{y}^{\frac{1}{2}}S_{x}^{\frac{1}{2}} \left[(18 \cdot 2^{n+2}-18)x^2y^2+(36 \cdot 2^{n+1}-24)x^2y^3+ (9 \cdot 2^{n+2}-9)x^3y^3+ 6x^3y^6 \right] \\
= & D_{x}^{\frac{1}{2}}JS_{y}^{\frac{1}{2}} \left[\frac{1}{\sqrt{2}} (18 \cdot 2^{n+2}-18)x^2y^2+\frac{1}{\sqrt{2}} (36 \cdot 2^{n+1}-24)x^2y^3+\frac{1}{\sqrt{3}} (9 \cdot 2^{n+2}-9)x^3y^3+\frac{1}{\sqrt{2}} (6x^3y^6) \right] \\
& = D_{x}^{\frac{1}{2}}J \left[\frac{1}{2} (18 \cdot 2^{n+2}-18)x^2y^2+\frac{1}{\sqrt{6}} (36 \cdot 2^{n+1}-24)x^2y^3+\frac{1}{3} (9 \cdot 2^{n+2}-9)x^3y^3+\frac{1}{\sqrt{18}} (6x^3y^6) \right] \\
& = D_{x}^{\frac{1}{2}}\left[\frac{1}{2} (18 \cdot 2^{n+2}-18)x^4+\frac{1}{\sqrt{6}} (36 \cdot 2^{n+1}-24)x^5+\frac{1}{3} (9 \cdot 2^{n+2}-9)x^6+\frac{1}{\sqrt{18}} (6x^9) \right] \\
& = (18 \cdot 2^{n+2}-18)x^4+\frac{\sqrt{5}}{\sqrt{6}} (36 \cdot 2^{n+1}-24)x^5+\frac{\sqrt{6}}{3} (9 \cdot 2^{n+2}-9)x^6+\frac{6}{\sqrt{2}} x^9.
\end{align*}

\begin{align*}
(iii) \, & S_{x}^{\frac{1}{2}}JD_{y}^{\frac{1}{2}}D_{x}^{\frac{1}{2}}(M(\Upsilon; x,y))  \\
&=S_{x}^{\frac{1}{2}}JD_{y}^{\frac{1}{2}}D_{x}^{\frac{1}{2}} \left[(18 \cdot 2^{n+2}-18)x^2y^2+(36 \cdot 2^{n+1}-24)x^2y^3+ (9 \cdot 2^{n+2}-9)x^3y^3+ 6x^3y^6 \right] \\
&=S_{x}^{\frac{1}{2}}JD_{y}^{\frac{1}{2}} \left[\sqrt{2}(18 \cdot 2^{n+2}-18)x^2y^2+\sqrt{2}(36 \cdot 2^{n+1}-24)x^2y^3+\sqrt{3} (9 \cdot 2^{n+2}-9)x^3y^3+ 6\sqrt{3}x^3y^6 \right] \\
&=S_{x}^{\frac{1}{2}}J \left[2(18 \cdot 2^{n+2}-18)x^2y^2+\sqrt{6}(36 \cdot 2^{n+1}-24)x^2y^3+3(9 \cdot 2^{n+2}-9)x^3y^3+ 6\sqrt{18}x^3y^6 \right] \\
&=S_{x}^{\frac{1}{2}} \left[(36 \cdot 2^{n+2}-36)x^4+\sqrt{6}(36 \cdot 2^{n+1}-24)x^5+(27 \cdot 2^{n+2}-27)x^6+ 6\sqrt{18}x^9 \right] \\
& =\frac{1}{2}(36 \cdot 2^{n+2}-36)x^4+\frac{\sqrt{6}}{\sqrt{5}}(36 \cdot 2^{n+1}-24)x^5+\frac{1}{\sqrt{6}}(27 \cdot 2^{n+2}-27)x^6+ \frac{6}{\sqrt{9}}\sqrt{18}x^9\\
& =(18 \cdot 2^{n+2}-18)x^4+\frac{\sqrt{6}}{\sqrt{5}}(36 \cdot 2^{n+1}-24)x^5+\frac{27}{\sqrt{6}}(2^{n+2}-1)x^6+ 6\sqrt{2}x^9. 
\end{align*}

Hence, the Nirmala indices of $\Upsilon$ are given by 

\begin{align*}
(a) \, N(\Upsilon)=& D_{x}^{\frac{1}{2}}J(M(\Upsilon; x,y))|_{x=1} \\
& = (36 \cdot 2^{n+2}-36)+\sqrt5(36 \cdot 2^{n+1}-24)+\sqrt6(9 \cdot 2^{n+2}-9)+18.
\end{align*} 
\begin{align*}
(b) \, IN_{1}(\Upsilon)=&D_{x}^{\frac{1}{2}}JS_{y}^{\frac{1}{2}}S_{x}^{\frac{1}{2}}(M(\Upsilon; x,y))|_{x=1}   \\
 & = (18 \cdot 2^{n+2}-18)+\frac{\sqrt{5}}{\sqrt{6}}(36 \cdot 2^{n+1}-24)+\frac{\sqrt{6}}{3}(9\cdot 2^{n+2}-9)+\frac{6}{\sqrt{2}}.
 \end{align*}
 \begin{align*}     
(c) \, IN_{2}(\Upsilon)=& S_{x}^{\frac{1}{2}}JD_{y}^{\frac{1}{2}}D_{x}^{\frac{1}{2}}(M(\Upsilon; x,y))|_{x=1}   \\
&=(18 \cdot 2^{n+2}-18)+\frac{\sqrt{6}}{\sqrt{5}}(36 \cdot 2^{n+1}-24)+\frac{27}{\sqrt{6}}(2^{n+2}-1)+6\sqrt{2}.
\end{align*}  
\subsection{Entropy measures of the complex structure of ruthenium bipyridine $RB_n$}
We proceed with the computation of graph entropy metrics for the complex structure of ruthenium bipyridine $RB_n$ using Shannon's entropy. Initially, we assess the degree-based graph entropy expression. Subsequently, we determine the mathematical formulations for entropy measures based on Nirmala indices, utilizing the previously derived expressions of the Nirmala indices.\\\\
\textbf{Nirmala entropy of $RB_n$:}\\
From Theorem 3.2, the  Nirmala index of $\Upsilon$ is given by
\begin{equation*}
N(\Upsilon)=(36 \cdot 2^{n+2}-36)+\sqrt5(36 \cdot 2^{n+1}-24)+\sqrt6(9 \cdot 2^{n+2}-9)+18.     
\end{equation*}
From Table 2 and Equation (7), we have
\begin{align*}
 ENT_{N}(\Upsilon)=&log(N(\Upsilon))  - \frac{1}{N(\Upsilon)}              \displaystyle \sum_{st \in E(\Upsilon)} \sqrt{d_{\Upsilon}(s)+d_{\Upsilon}(t)} \times log (\sqrt{d_{\Upsilon}(s)+d_{\Upsilon}(t)}) \\
 = & log(N(\Upsilon))-\frac{1}{N(\Upsilon)} \left[\displaystyle \sum_{i=1}^{4} \displaystyle \sum_{st \in E_{i}(\Upsilon)} \sqrt{d_{\Upsilon}(s)+d_{\Upsilon}(t)} \times log (\sqrt{d_{\Upsilon}(s)+d_{\Upsilon}(t)}) \right] \\
 & = log(N(\Upsilon))-\frac{1}{N(\Upsilon)} \left[(18 \cdot 2^{n+2}-18) \cdot \sqrt{2+2} \cdot log(\sqrt{2+2}) \right] \\
 & -\frac{1}{N(\Upsilon)} \left[(36 \cdot 2^{n+1}-24) \cdot \sqrt{2+3} \cdot log(\sqrt{2+3}) \right] \\
 - &\frac{1}{N(\Upsilon)} \left[(9 \cdot 2^{n+2}-9) \cdot \sqrt{3+3} \cdot log(\sqrt{3+3}) + (6 \cdot \sqrt{3+6} \cdot log(\sqrt{3+6})\right] \\
  & = log(N(\Upsilon))-\frac{1}{N(\Upsilon)} \left[(18 \cdot 2^{n+2}-18) \cdot 2 \cdot log(\sqrt{2+2}) \right] \\
 & -\frac{1}{N(\Upsilon)} \left[(36 \cdot 2^{n+1}-24) \cdot \sqrt{5} \cdot log(\sqrt{5}) \right] \\
 - &\frac{1}{N(\Upsilon)} \left[(9 \cdot 2^{n+2}-9) \cdot \sqrt{6} \cdot log(\sqrt{6}) + 6 \cdot 3 \cdot log(3) \right]  \\
 & = log(N(\Upsilon))-\frac{1}{N(\Upsilon)} \left[(18 \cdot 2^{n+2}-18) \cdot 2 \cdot log(\sqrt{2+2}) \right] \\
 & -\frac{1}{N(\Upsilon)} \left[(36 \cdot 2^{n+1}-24) \cdot \sqrt{5} \cdot log(\sqrt{5}) \right] \\
 - &\frac{1}{N(\Upsilon)} \left[(9 \cdot 2^{n+2}-9) \cdot \sqrt{6} \cdot log(\sqrt{6}) + 18 \cdot log(3) \right] 
  \end{align*}
Finally, we get the desired formulation of the Nirmala entropy for $RB_n$ by substituting the value of $N(\Upsilon)$ into the previous expression.\\\\
\textbf{First inverse Nirmala entropy of $RB_{n}$:}\\
From Theorem 3.2, the first inverse Nirmala index of $\Upsilon$ is given by
\begin{align*}
IN_{1}(\Upsilon)=& (18 \cdot 2^{n+2}-18)+\frac{\sqrt{5}}{\sqrt{6}}(36 \cdot 2^{n+1}-24)+\frac{\sqrt{6}}{3}(9\cdot 2^{n+2}-9)+\frac{6}{\sqrt{2}}.    
\end{align*}
From Table 2 and Equation (8), we have
\begin{align*}
 ENT_{IN_{1}}(\Upsilon)=&log(IN_{1}(\Upsilon))  - \frac{1}{IN_{1}(\Upsilon)}              \displaystyle \sum_{st \in E(\Upsilon)} \sqrt{\frac{1}{d_{\Upsilon}(s)}+\frac{1}{d_{\Upsilon}(t)}} \times log \left( \sqrt{\frac{1}{d_{\Upsilon}(s)}+\frac{1}{d_{\Upsilon}(t)}} \right) \\
 =& log(IN_{1}(\Upsilon)) - \frac{1}{IN_{1}(\Upsilon)} \left[\displaystyle \sum_{i=1}^{4} \displaystyle \sum_{st \in E_{i}(\Upsilon)} \sqrt{\frac{1}{d_{\Upsilon}(s)}+\frac{1}{d_{\Upsilon}(t)}} \times log \left( \sqrt{\frac{1}{d_{\Upsilon}(s)}+\frac{1}{d_{\Upsilon}(t)}} \right) \right] \\
 = & log(IN_{1}(\Upsilon)) - \frac{1}{IN_{1}(\Upsilon)} \left[(18 \cdot 2^{n+2}-18) \cdot \sqrt{\frac{1}{2}+\frac{1}{2}} \cdot log\left(\sqrt{\frac{1}{2}+\frac{1}{2}}\right) \right] \\
 - & \frac{1}{IN_{1}(\Upsilon)} \left[(36 \cdot 2^{n+1}-24) \cdot \sqrt{\frac{1}{2}+\frac{1}{3}} \cdot log\left(\sqrt{\frac{1}{2}+\frac{1}{3}}\right) \right] \\
 - & \frac{1}{IN_{1}(\Upsilon)} \left[(9 \cdot 2^{n+2}-9) \cdot \sqrt{\frac{1}{3}+\frac{1}{3}} \cdot log\left(\sqrt{\frac{1}{3}+\frac{1}{3}}\right) \right] \\
 - & \frac{1}{IN_{1}(\Upsilon)} \left[6 \cdot \sqrt{\frac{1}{3}+\frac{1}{6}} \cdot log\left(\sqrt{\frac{1}{3}+\frac{1}{6}}\right) \right]
 \end{align*}
 Since $log(1)=0$, 
 \begin{align*}   
 ENT_{IN_{1}}(\Upsilon)= & log(IN_{1}(\Upsilon))  
 - \frac{1}{IN_{1}(\Upsilon)} \left[(36 \cdot 2^{n+1}-24) \cdot \sqrt{\frac{5}{6}} \cdot log\left(\frac{5}{6}\right) \right] \\
 - & \frac{1}{IN_{1}(\Upsilon)} \left[ (9 \cdot 2^{n+2}-9) \cdot \sqrt{\frac{2}{3}}  \cdot log\left(\sqrt{\frac{2}{3}}\right) + 6 \cdot \sqrt{\frac{1}{2}} \cdot log\left(\sqrt{\frac{1}{2}}\right) \right]  
\end{align*}
Finally, by substituting the value of $IN_{1}(\Upsilon)$ into the preceding expression, we obtain the desired formulation of the first inverse Nirmala entropy for $RB_n$.\\\\
\textbf{Second inverse Nirmala entropy of $RB_n$:}\\
From Theorem 3.2, the second inverse Nirmala index of $\Upsilon$ is given by
\begin{align*}
IN_{2}(\Upsilon)=&(18 \cdot 2^{n+2}-18)+\frac{\sqrt{6}}{\sqrt{5}}(36 \cdot 2^{n+1}-24)+\frac{27}{\sqrt{6}}(2^{n+2}-1)+6\sqrt{2}.
\end{align*} 
From Table 3 and Equation (9), we have
\begin{align*}
 ENT_{IN_{2}}(\Upsilon)=& log(IN_{2}(\Upsilon))  - \frac{1}{IN_{2}(\Upsilon)}              \displaystyle \sum_{st \in E(\Upsilon)} \frac{\sqrt{d_{\Upsilon}(s) \cdot  d_{\Upsilon}(t)} }{\sqrt{d_{\Upsilon}(s) + d_{\Upsilon}(t)} } \times log \left( \frac{\sqrt{d_{\Upsilon}(s) \cdot  d_{\Upsilon}(t)} }{\sqrt{d_{\Upsilon}(s) + d_{\Upsilon}(t)} } \right) \\
 = & log(IN_{2}(\Upsilon))  - \frac{1}{IN_{2}(\Upsilon)} \left[ \displaystyle \sum_{i=1}^{4} \displaystyle \sum_{st \in E_{i}(\Upsilon)}              \frac{\sqrt{d_{\Upsilon}(s) \cdot  d_{\Upsilon}(t)} }{\sqrt{d_{\Upsilon}(s) + d_{\Upsilon}(t)} } \times log \left( \frac{\sqrt{d_{\Upsilon}(s) \cdot  d_{\Upsilon}(t)} }{\sqrt{d_{\Upsilon}(s) + d_{\Upsilon}(t)} } \right) \right] \\
 = & log(IN_{2}(\Upsilon))  - \frac{1}{IN_{2}(\Upsilon)} \left[ (18 \cdot 2^{n+2}-18) \cdot \frac{\sqrt{4}}{\sqrt{4}} \cdot log  \left(\frac{\sqrt{4}}{\sqrt{4}} \right) \right]\\ 
  & - \frac{1}{IN_{2}(\Upsilon)} \left[ (36 \cdot 2^{n+1}-24) \cdot \frac{\sqrt{6}}{\sqrt{5}} \cdot log  \left(\frac{\sqrt{6}}{\sqrt{5}} \right) \right]\\
 - & \frac{1}{IN_{2}(\Upsilon)} \left[(9 \cdot 2^{n+2}-9) \cdot \frac{\sqrt{9}}{\sqrt{6}} \cdot log \left(\frac{\sqrt{9}}{\sqrt{6}} \right) +6 \cdot \frac{\sqrt{18}}{\sqrt{9}} \cdot log \left(\frac{\sqrt{18}}{\sqrt{9}} \right) \right] 
 \end{align*} 
Since $log(1)=0$, 
\begin{align*}
ENT_{IN_{2}}(\Upsilon)=& log(IN_{2}(\Upsilon))  - \frac{1}{IN_{2}(\Upsilon)} \left[ (36 \cdot 2^{n+1}-24) \cdot \frac{\sqrt{6}}{\sqrt{5}} \cdot log  \left(\frac{\sqrt{6}}{\sqrt{5}} \right) \right]\\
 - & \frac{1}{IN_{2}(\Upsilon)} \left[(9 \cdot 2^{n+2}-9) \cdot \frac{3}{\sqrt{6}} \cdot log \left(\frac{3}{\sqrt{6}} \right) + 6 \cdot \sqrt{2} \cdot log \left(\sqrt{2} \right) \right] 
\end{align*}
The second inverse Nirmala entropy for $RB_n$ can finally be expressed as desired by substituting the value of $IN_{2}(\Upsilon)$ in the previous expression.

\section{Comparison through numerical demonstrations}
Numerous scientific fields, such as computer science, information theory, chemistry, biological therapies, and pharmacology, frequently use graph entropy measurements. To accurately quantify these molecular properties, scientists working in these disciplines rely on numerical calculation representation. This section uses numerical computation to compare the Nirmala indices and the accompanying entropy measures. The numerical computation of the Nirmala indices and accompanying entropy measures for $RB_{n}$, where $1\leq n \leq 25$ is presented in Table 3. \\
\textbf{Table 3}. Calculated values of the Nirmala indices and their associated entropy measures of $RB_{n}$, where $1 \leq n \leq 25$.  
\newpage
\begin{center} 
\begin{table}[h!]
\centering
\renewcommand{\arraystretch}{1.70}
\begin{tabular}{|>{\centering\arraybackslash}m{0.5cm}||>{\centering\arraybackslash}m{2.8cm} |>{\centering\arraybackslash}m{2.8cm} |>{\centering\arraybackslash}m{2.8cm} |>{\centering\arraybackslash}m{1.6cm} |>{\centering\arraybackslash}m{1.6cm} |>{\centering\arraybackslash}m{1.6cm}  ||} 
 \hline
 [n] & N & $IN_{1}$ & $IN_{2}$ & $ENT_{N}$ &  $ENT_{IN_{1}}$  &   $ENT_{IN_{2}}$
 \\ [0.5ex] 
 \hline\hline
[1] & 692.64 & 291.22 & 343.09 & 5.7484 & 5.7292 & 5.7489 \\
\hline
[2] & 1479.00 & 625.46 & 733.02 & 6.5111 & 6.5117 & 6.5114  \\ 
\hline
[3] &  3051.71 & 1293.94 & 1512.90 & 7.2373 & 7.2377 & 7.2375\\ 
 \hline 
[4] & 6197.14 & 2630.91 & 3072.60 & 7.9466 & 7.9469 & 7.9468\\ 
\hline 
[5] & 12488 & 5304.84 & 6192 & 8.6477 & 8.6480 & 8.6479 \\
\hline 
[6] & 25069.71 & 10652.70 & 12430.79 & 9.3448 & 9.3451 &9.3450 \\
\hline
[7] & 50233.14 & 21348.41 & 24908.40 & 10.0400 & 10.0402 & 10.0401 \\
\hline
[8] & 100559.99 & 42739.85 & 49863.66 & 10.7341 & 10.7344 & 10.7342 \\
\hline 
[9] & 201213.69 & 85500.72 & 99774.15 &11.4277 & 11.4280 & 11.4279\\
\hline  
[10] & 402521.09 & 171088.46 & 199595.14 & 12.1211 & 12.1214 & 12.1213\\
\hline
[11] & 805135.90 & 342219.94 & 399237.11 & 12.8144 & 12.8147 & 12.8145 \\
\hline 
[12] & 1610365.52 & 684482.89 & 798521.06 & 13.5076 & 13.5079 & 13.5077 \\
\hline 
[13] & 3220824.76 & 1369008.80 & 1597088.95 & 14.2008 & 14.2011 & 14.2009 \\
\hline 
[14] & 6441743.24 & 2738060.63 & 3194224.72 & 14.8940 & 14.8942 & 14.8941 \\
\hline
[15] & 12883580.20 & 5476164.27 & 6388496.28 & 15.5871 & 15.5874 & 15.5872 \\
\hline
[16] & 25767254.11 & 10952371.56 & 12777039.40 & 16.2803 & 16.2805 & 16.2804 \\
\hline 
[17] & 51534601.93 & 21904786.14 & 25554125.63 & 16.9734 & 16.9737 & 16.9735 \\
\hline 
[18] & 103069297.57 & 43809615.30 & 51108298.10 & 17.6666 & 17.6668 & 17.6667 \\
\hline 
[19] & 206138688.86 & 87619273.63 & 102216643.02 &18.3597 &18.3600 &18.3598 \\
\hline 
[20] & 412277471.44 & 175238590.27 & 204433332.88 & 19.0529 & 19.0531 & 19.0530 \\
\hline
[21] & 82455036.60 & 350477223.57 & 408866712.60 & 19.7460 & 19.7463  &19.7461\\
\hline
[22] & 1649110166.91 & 700954490.15 & 817733472.03 & 20.4391 & 20.4394 & 20.4393 \\
\hline
[23] & 3298220427.54 & 1401909023.33 & 1635466990.89 & 21.1323 & 21.1326 & 21.1324  \\
\hline
[24] & 6596440948.79 & 2803818089.67 & 3270934028.62 & 21.8254 & 21.8257 & 21.8256\\
\hline
[25] & 13192881991.29 & 5607636222.37 & 6541868104.06 & 22.5186 & 22.5189 & 22.5187 \\
\hline
\end{tabular}
\end{table} 
\end{center}
From Table 3, the following two remarks are possible. \\
\textbf{Remark 1:}  The Nirmala indices and associated entropy measures of the complex structure of ruthenium bipyridine increase as the values of $n$
increase. \\
\textbf{Remark 2:} For complex structure of ruthenium bipyridine graph $\Upsilon=RB_{n}$, we have the following inequality relationships: 
\begin{center}
$IN_1(\Upsilon) < IN_2(\Upsilon) < N(\Upsilon)$ 
\end{center}
\begin{center}
$ENT_{N}(\Upsilon) \approx ENT_{IN_{1}}(\Upsilon) \approx ENT_{IN_{2}}(\Upsilon)$.
\end{center}
\subsection{The logarithmic regression model}
To investigate the relationship between the dependent variable and one or more predictor variables in our dataset, we apply logarithmic regression analysis. Logarithmic regression is a nonlinear technique that modifies the dependent or predictor variables using logarithmic functions. The following is the equation for the logarithmic regression model.
\begin{equation*}
y=a*log(x)+b, 
\end{equation*} 
where the response variable is $y$ and the predictor variable is $x$. The regression coefficients $a$ and $b$ represent the relation between $x$ and $y$.

Here, for $1 \leq n \leq 25$, we examine the link between the Nirmala indices and entropy metrics of the complex structure of ruthenium bipyridine $RB_{n}$ using a logarithmic regression analysis. The study utilized many statistical measurements, such as the squared correlation coefficient ($R^{2}$), the sum of square error (SSE), adjusted squared correlation coefficient (Adj. R-sq), root mean square error (RMSE), and squared correlation coefficient ($R^{2}$). A low RMSE value (nearer to 0) implies that the model performs well, whereas a larger $R^{2}$ value (near 1) suggests that the regression line fits the data better. In this instance, obtaining a larger $R^{2}$ value is our main goal. 

The statistics of curve fitting of the Nirmala indices versus Nirmala entropy measures for the complex structure of ruthenium bipyridine $RB_n$ using the logarithmic regression are shown in Table 4.
\newpage
\vspace{5mm}
\begin{table}[h!]
\centering
\renewcommand{\arraystretch}{2.5}
\begin{tabular}{|>{\centering\arraybackslash}m{7cm}||>{\centering\arraybackslash}m{1.8cm} |>{\centering\arraybackslash}m{1.8cm} |>{\centering\arraybackslash}m{2cm} |>{\centering\arraybackslash}m{2cm} ||} 
 
 \hline
 Model & $R^{2}$ & SSE & Adj. R-sq & RMSE
 \\ [0.5ex] 
 \hline\hline
 $ENT_{N}=1.0129*log(N)-0.8842$  &  \textbf{0.9920}   & 4.9922 & 0.9917 & 0.4658 \\
\hline
 $ENT_{IN_{1}}=1.0002*log(IN_{1})+0.0683$  & \textbf{1}   & 0.000242 & 1 & 0.0032  \\
\hline
 $ENT_{IN_{2}}=1.0002*log(IN_{2})+0.0854$  & \textbf{1}    & 0.000316 & 1 & 0.0011   \\
 \hline
\end{tabular}
\end{table} 
  
\textbf{Table 4}. Statistics of curve fitting of the Nirmala indices vs. Nirmala entropy measures of the complex structure of ruthenium bipyridine $RB_n$.
\vspace{15mm}
\begin{figure}[h!]
\centering
\includegraphics[width=140mm]{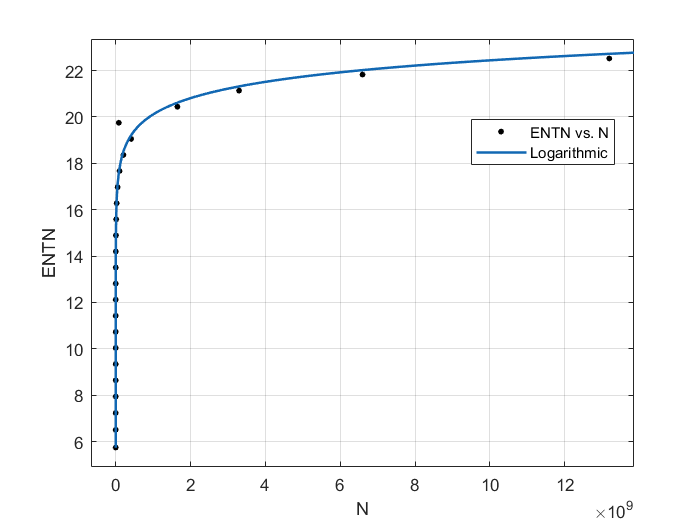}  
  \end{figure} 
  \newpage
  \begin{figure}[h!]
\centering
\includegraphics[width=140mm]{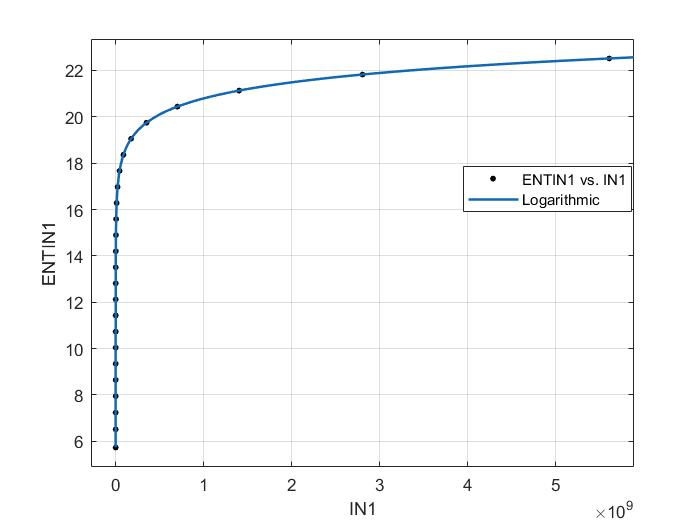}  
  \end{figure} 
 \begin{figure}[h!]
\centering
\includegraphics[width=140mm]{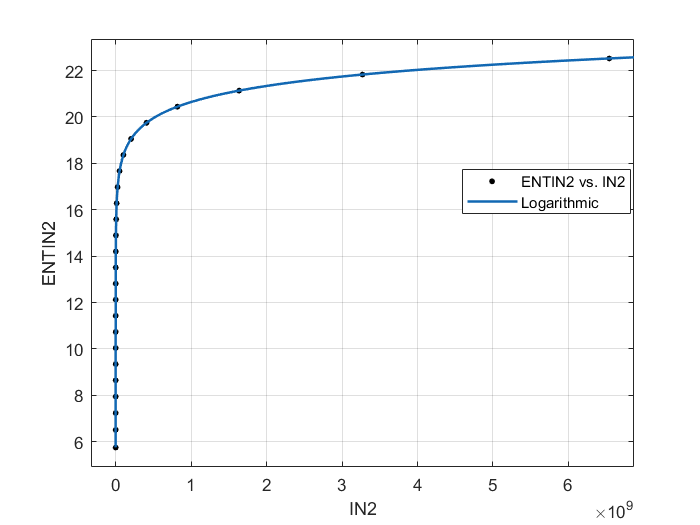}  
  \end{figure} 
  \textbf{Figure 3}. Curve fitting plots for the Nirmala indices vs. Nirmala entropy measures of the complex structure of ruthenium bipyridine $RB_n$.

\newpage
 \section{Discussion} 
Researchers can gain a better understanding of the molecular properties and behavior of chemical and biological systems by using topological indices, which are numerical representations of molecular groups in graph theory. Accurate calculation of numerical indices gives scholars relevant information that enhances their knowledge of the topic. In this work, we study the so-called degree-based topological indices, that is, Nirmala indices, of the complex structure of ruthenium bipyridine $RB_n$. Table 3 shows that as $n$ increases, the Nirmala indices and related entropy measures of $RB_n$ also increase. Entropy metrics assess the uncertainty or information content of a dataset to assist estimate its complexity and distribution. Data analysis, thermodynamics, and information theory heavily rely on these concepts. Accurate numerical entropy calculations provide academics with valuable insights, enhancing their comprehension of the network under investigation. Considering its advantages, it validates our focus on the edge weight entropy of $RB_n$. Table 3 illustrates that entropy measurements of the complex structure of ruthenium bipyridine $RB_n$ rise with increasing $n$.
 
To account for non-linearity in data and produce predictions, the social sciences, biology, and economics employ logarithmic regression, which is a non-linear regression technique. The study's statistical measures, such as the squared correlation coefficient ($R^{2}$), root mean square error (RMSE), adjusted squared correlation coefficient (Adj. R-sq), and the sum of square errors (SSE), are shown in Table 4. A regression line that fits the data better is indicated by a greater $R^{2}$ value, which is closer to 1. The Nirmala indices and the related entropy measure values of $RB_{n}$ appear to fit the curve well, as shown in Figure 3.

\section{Conclusion}
This research has made use of the definitions of the Nirmala indices and entropy measures derived from them. A mathematical formulation of the Nirmala indices of the complex structure of ruthenium bipyridine $RB_{n}$ has been achieved. The entropy measures of $RB_{n}$ based on the Nirmala indices have been analyzed using its M-polynomial. Table 3 shows that when $n$ increases, the Nirmala indices and related entropy measurements of the complex structure of ruthenium bipyridine $RB_{n}$ also increase. Additionally, the inequality relationships shown below are also accurate. 
\begin{center}
$IN_1(\Upsilon) < IN_2(\Upsilon) < N(\Upsilon)$ 
\end{center}
\begin{center}
$ENT_{N}(\Upsilon) \approx ENT_{IN_{1}}(\Upsilon) \approx ENT_{IN_{2}}(\Upsilon)$.
\end{center}

The topology and structural characteristics of the complex structure of ruthenium bipyridine $RB_{n}$ will be examined with the help of the study's results.

\section*{Funding} No funding is available for this study.
\section*{Author contributions} All authors contributed equally.
\section*{Data Availability Statement}
This manuscript has no associated data.
\section*{Declarations}
\textbf{Conflict of interest} The authors declare that they have no known competing financial interests or personal relationships that could have appeared to influence the work reported in this paper.

  \makeatletter
\renewcommand{\@biblabel}[1]{[#1]\hfill}

\makeatother

\end{document}